# Study of Memristor-based Oscillatory Neural Networks using PPV modeling

Hanyu Wang, Miao Qi, Bo Wang, *Member, IEEE*

*Abstract*—Memristor-based oscillator is becoming promising thanks to its inherent NDR (Negative Differential Region) property and compact circuit structure. This paves the way to the large scale oscillatory neural network (ONN) and the realization of pattern recognition based on its global synchronization. However, the simulation of large scale ONN encounters the problem of long simulation time because of the large number of oscillators. Here we propose a highly efficient method to abstract the phase sensitivity characteristic of the memristor-based oscillator, i.e., its PPV (Perturbation Projection Vector), which allows reducing considerably the complexity of ONN simulation, and speeding up the simulation more than 2000 times. Our study also reveals the impact of the circuit parameters on the pattern recognition accuracy and the robustness against the frequency mismatch.

## I. Introduction

Oscillators exist ubiquitously in nature and electronic field. Many oscillators can be coupled together to construct oscillation array. If synchronized, these oscillators are able to achieve the pattern recognition of pixels [1]. For such tasks, a large amount of oscillators is required (e.g., 30*40 pixels need 1200 oscillators) [1], which brings challenges in designing small footprint oscillator as well as in fast circuit analysis.

[2] employed MOSFET-based LC oscillator as building block of array, which ignores the large chip area of inductors. [3] adopted the area-efficient memristor-based oscillators, and incorporated them in PLLs. If simulated at transistor level, it becomes very time-consuming (several hours even days) due to the large number of the oscillators and the voltage/current details in each oscillator circuit.

In this paper, we propose a highly efficient method to quickly simulate the large scale ONN, which is based on the PPV modeling of the memristor-based oscillator. This allows reducing the simulation time to several seconds. We first abstract the PPV of the oscillator, which models efficiently its phase behaviors. Then we build two coupled oscillators and the ONN with the proposed inverter-based coupling circuit. The simulation of ONN synchronization is instantly accomplished with the correct pattern recognition. In addition, we find that sinusoidal waveform shows better recognition accuracy than the sawtooth waveform. This can help guide the design of memristor-based oscillator. Also we have studied the effect of group frequency deviation on the synchronization.

This paper is organized as follows: in section II, we study the characteristic of memristor-based oscillator. Section III presents the PPV modeling of memristor-based oscillator. Section IV describes the coupler circuit in ONN. Finally, in Section V and VI, we illustrate the architecture of ONN and the numerical experiments for pattern recognition using PPV.

## II. Memristor-Based Oscillator

We first introduce the memristor and its NDR property. Then, we simulate the memristor-based oscillator based on the existing unfolding polynomial method.

## A. Negative Differential Resistance

To sustain a stable oscillation, an equivalent negative resistance is necessary to compensate the energy dissipated by passive resistance. Some memristor devices have demonstrated the negative resistance characteristic (NDR), which can be identified in I-V curve, as is shown in Fig.1 [4].

It's easy to see that below the turning point A, the memristor remains OFF state; between point A and point B, the device turns from insulator phase to metal phase with voltage drop and current increase, implying the negative resistance (NDR).

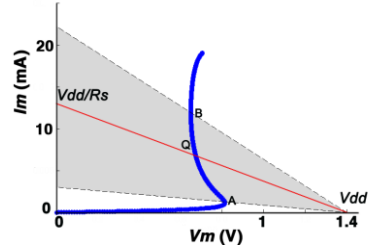

Fig. 1 Negative Differential Resistance(NRD) curve of the memristor

## B. Oscillator Circuit

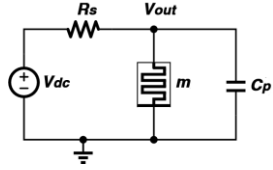

Fig.2 Memristor-based Oscillator Circuit

An oscillator circuit based on the NDR of the memristor is illustrated in Fig.2. When memristor is in OFF state, the capacitor $C_p$ is charged through $R_s$; Once the voltage of $C_p$ is large enough, the memristor turns on, entering the NDR zone with voltage drop and current increase, $C_p$ then gets discharged. The repeat of the charge/discharge forms the oscillation.

By carefully choosing the serial resistance and source voltage, we can have a load line with a unique operating point Q in NDR region (shown in Fig.1) which allows for the oscillation [4].

## C. Model for Oscillator

The key of modeling the oscillator circuit relies on the accurate characterization of the memristor. As known, the memristor can be described in relation between electric charge and magnetic flux leakage [5]. This implies that it can be modeled generally by differential equation regardless of its material. [6] developed a memristor model by using unfolding polynomial, which consists of two equations: (1) A differential equation of state variable $x$ in the form of polynomial. (2) An algebra equation relating the voltage and the current of the memristor (i.e., conductance) is in the form of polynomial too.

$$\frac{dx}{dt} = f(x, v_m) = \sum_{i=0}^{1} a_i x^i + b_2 v_m^2 + \sum_{i=1}^{5} c_{2i} v_m^2 x^i \qquad (1)$$

$$i_m = g(x) v_m = v_m \sum_{i=0}^{5} d_i x^i \qquad (2)$$

In this paper, we adopt the same polynomial coefficients values as those in [4], since they have been carefully validated by experiments.

The oscillator described by the above equations is simulated in MATLAB, as is shown in Fig.3, with $V_{dc}$=3.3V, $R_s$=810Ohm, $C_p$=800pF. The equations are also implemented in Verilog-A using Euler method and simulated on Cadence. As shown in Fig.3(a), their results match very well.

Compared with the physical model with internal detailed working mechanism of the memristor, this unfolding polynomial model is much simpler and efficient, and works fine for one oscillator. Nevertheless, the simulation of ONN with large number of oscillators would become very time-consuming, as will be seen in section VI, because these equations includes the details of voltages and current of devices and their higher nonlinear harmonics, which considerably slows down the simulation. Fortunately, the amplitude of voltage/current has little influence on the oscillator coupling and can be omitted without harm. So only the phase is sufficient to describe the oscillator [7][8]. In the next section, we will use PPV to model the phase of the oscillator.

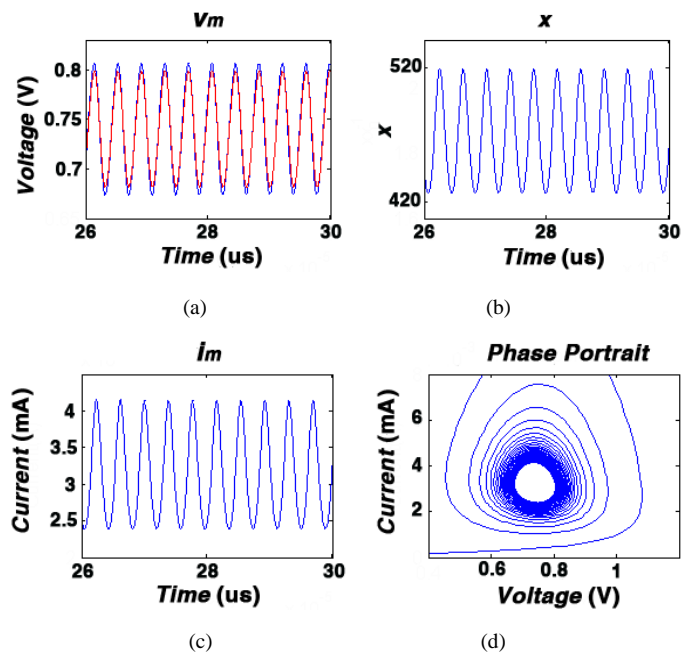

(a)    (b)

(c)    (d)

Fig.3 Oscillator simulation: (a) Voltage output: MATLAB(blue line) and Cadence(red line) (b)State variable (c)Current (d)Limit cycle

### III. MODELING OF MEMRISTOR OSCILLATOR USING PPV

In the ONN, what interests us most is the phase of the oscillator and its phase response to the injection signal. PPV (Perturbation Projection Vector) are means that quantitatively describe phase sensitivity of an oscillator.

PPV is initially proposed and used in electronic oscillators [8]. It defines degree of phase sensitivity to the injected signal and can be regarded as phase transfer function of small inputs. PPV is unique for one oscillator, so it's a good choice for describing the phase behavior of memristor-based oscillator.

However, one often encounters difficulties while abstracting the PPV. A precise and efficient PPV calculation method is to employ PSS and PXF analysis in Cadence [10]. However, almost all the memristor (physical model or behavioral model) are described by Verilog-A, in which the hidden-state prevent the PSS analysis from convergence. So we cannot directly obtain the PPV of the memristor-based oscillator. Fortunately, we can first calculate its PRC, and then transform it into PPV.

As for PRC, it's initially proposed for biological oscillators, which also describes the phase shift of an oscillator under external perturbation (light, current, etc.). However, PRC varies with the amplitude of perturbation, so it's not unique. The calculation of PRC is easy with transient simulation [9].

Inspired by this, we abstract the PRC by injecting a series of pulse current into the $V_{out}$ terminal of the oscillator, then measure the phase shift after 20 oscillation periods. It can be quickly achieved within several minutes by OCEAN script. Then through the method proposed in [11], PRC can be converted into PPV by using:

$$\Gamma(t) = \frac{P(t)}{b \cdot h \cdot \omega_0} \tag{3}$$

where Γ(t) is PPV function and P(t) is PRC function. $b$ and $h$ are height and width of pulse current respectively. $\omega_0$ is the angular frequency in oscillation.

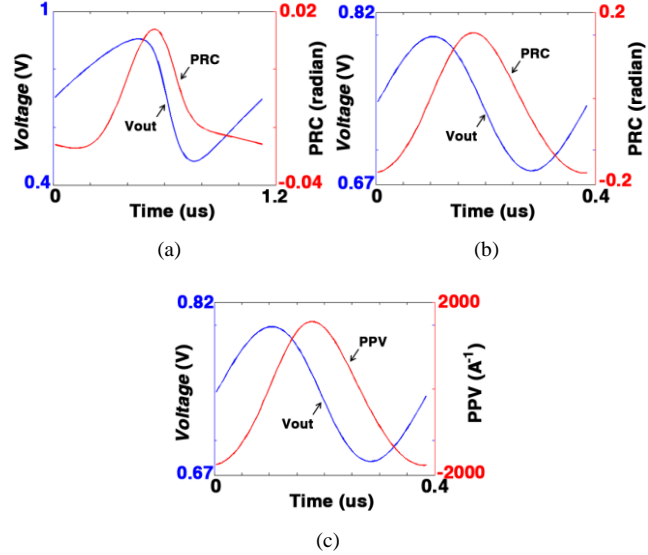

Fig.4 Voltage output(blue line) and PRC/PPV (red line) (a) *Rs*=1k Ohm, *Cp*=3500pF (b) PRC *Rs*=810 Ohm, *Cp*=800pF (c) PPV of (b) [11]

Fig.4(a) shows the PRC curve of memristor-based oscillator, with the pulse current of 1mA and 6ns of width. The voltage output curve looks like sawtooth, which means that charging is slow while discharging is fast. Accordingly, the PRC curve demonstrates an asymmetric "V" shape. This implies that the output signal contains high-order harmonics, which will bring difficulties for the circuit analysis and for the oscillator synchronization (see Section V and VI).

To simplify the curve and analysis, we manage to balance the ratio of charging/discharging time by decreasing the value of capacitor and series resistance to 800 pF and 810 ohms, which allows decreasing the charging time. As is illustrated in Fig.4(b), the time of charging and discharging get to be nearly equal; also the positive and negative part of PRC become comparable and the waveform is almost of sinusoid.

As a result, both the voltage output and PRC are odd function, and they can easily turn into even function by a simple phase shift, thus satisfying the "odd-even" condition of ONN synchronization. [1]

After that, the PRC curve is transformed to PPV (Fig.4(c)). Once the PPV of memristor-based oscillator is obtained, it can be used to efficiently model the 2 coupled oscillators as well as large-scale ONN.

## IV. TWO COUPLED MEMRISTOR-BASED OSCILLATORS

We will construct the coupler circuit between oscillators and then simulate the two coupled memristor-based oscillators, since they are the basic blocks of ONN.

## A. Oscillatory Neurons

A distinction of this memristor-based circuit is that the voltage output and the injecting point share the same terminal. This enables each neuron circuit to couple with others using only one terminal, which greatly simplifies the interconnection of many oscillatory units in ONN.

## B. Digital Synapses (Coupler Circuit)

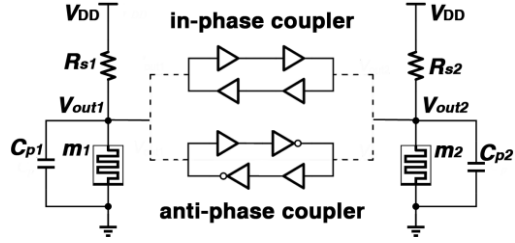

Fig.5 Two coupled oscillators with inverter-based couplers

The synapse in ONN is often realized by coupler circuit. In the neural networks, the amount of synapses is much more than that of neurons. This demands the coupler circuit to be area-efficient and low-power. In [2], the oscillators are coupled through differential pair transistors. In [12], an power-consuming amplifier is used as coupler circuit. As known, the above low-density analog circuits often occupy large chip areas and cannot scale down with the technology.

Here we propose a synapse (coupler) circuit based on digital inverter, which is very compact with good scalability. As is shown in Fig.5, the upper coupler circuit consists of two buffers and each buffer includes two inverters; thus input signal and output signal are in-phase. Another one includes a buffer and an inverter. Thus with odd number of inverters, the input and output are anti-phase. Therefore, these two sets of coupler circuit are able to realize respectively positive and negative transconductance. Besides, in pattern recognition the value of transconductance is integer multiple of a single coupler. Accordingly, by simply selecting the number of parallel couplers the transconductance can be adjusted readily.

Note that the injecting current of coupler circuit is determined by the driving capacity of the inverter. Here the output current of one inverter is set to be ±5.33uA.

## C. Verification of Coupled Oscillators

On one hand, we simulate the circuits in Cadence with oscillator modeled by unfolding polynomial [4] and coupler at transistor-level. By selecting in-phase or anti-phase coupling, the corresponding voltage outputs for in-phase or anti-phase synchronization are shown in Fig.6 (a) and (b).

On the other hand, the PPV model of bi-directional coupled oscillators can be represented in MATLAB as follows:

$$\dot{\alpha}_1(t) = \Gamma_1(t+\alpha_1(t)) \cdot I_{12} \cdot sign(V_{out2}(t+\alpha_2(t))) \qquad (4)$$

$$\dot{\alpha}_2(t) = \Gamma_2(t+\alpha_2(t)) \cdot I_{21} \cdot sign(V_{out1}(t+\alpha_1(t))) \qquad (5)$$

where $\alpha_1(t)$ and $\alpha_2(t)$ are time shifts caused by synchronization, $V_{out1}$ and $V_{out2}$ are output voltages of two oscillators, and $\Gamma_1(t)$ and $\Gamma_2(t)$ are the PPV functions corresponding to the output nodes. The sign function indicates the direction of the injected current. $I_{12}$ and $I_{21}$ are the values of output current in coupler circuits. We also assume that both oscillators are identical, with the same oscilltion frequency, amplitude and PPV function.

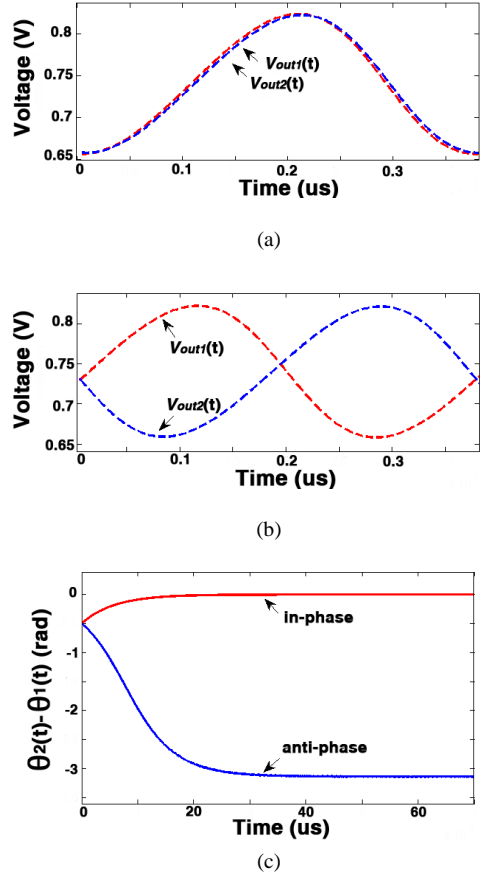

Fig.6 (a)In-phase and (b)anti-phase output voltage(Cadence) (c)Phase difference evolution(in-phase and anti-phase) (MATLAB), $\theta_1$ and $\theta_2$ are phase of oscillator 1 and 2.

The MATLAB simulation with PPV model is shown in Fig.6(c): the phase difference between two oscillators gradually reaches the steady values. The zero phase difference (red line) implies in-phase synchronization ($I_{12}=I_{21}=5.53uA$); another one in blue line is π and indicates anti-phase synchronization ($I_{12}=I_{21}=-5.53uA$). Such results match very well with the transient simulation in Cadence (Fig. 6(a,b)).

Now that we have verified the equivalence of the PPV model on MATLAB and circuit simulation on Cadence, we are ready to use PPV to evaluate the evolution of phase difference of ONN instead of using complex unfolding polynominal.

## V. Array Of Coupled Memristor Oscillator for Patten Recognition

In this section we extend from 2 coupled oscillators to the array of coupled oscillators. Fig.7 shows an example of ONN consisting of 5 globally connected oscillators (i.e., N=5). Based on PPV model of the oscillator, we will construct the model of ONN used for the pattern recognition.

According to the PPV modeling, the temporal evolution of time shift of $n^{th}$ oscillator can be described as the sum of the impact of all the other coupled oscillators [2]:

$$\dot{\alpha}_n(t) = \Gamma_n(t+\alpha_n(t)) \cdot \sum_{j=1}^{N} I_{nj} \cdot sign\left[V_{outj}(t+\alpha_j(t))\right] \tag{6}$$

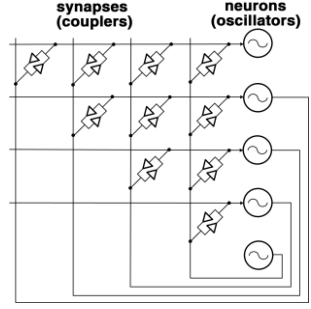

Fig.7 An Oscillatory Neural Network with 5 oscillators(neurons)

for $n=1,...,N$. $\omega_n$ is the angular frequency, $\alpha_n(t)$ is the time shift caused by synchronization; $I_{nj}$ is the injection current from the $j^{th}$ oscillator; $V_{outn}(t)$ and $\Gamma_n(t)$ are voltage and PPV function respectively. And Fourier transformation allows $\Gamma_n(t)$ to be presented as:

$$\Gamma_n(t) = \sum_{i=0}^{P}\left(A_i \cos(i\theta_n) + B_i \sin(i\theta_n)\right) \tag{7}$$

where $\theta_n = \omega_n(t + \alpha_n(t))$ is the total phase, $A_i$ and $B_i$ are Fourier coefficients, and $P$ determines the accuracy.

Also, the *sign* function can be unfolded as:

$$sign(V_{out,j}) \approx \sum_{i=1}^{Q}(-1)^{(i+1)}\frac{1}{2i-1}\cos(2i-1)\theta_j \tag{8}$$

where $Q$ determines approximation accuracy.

After averaging and transformation, the evolution of the total phase $\theta_n(t)$ of one oscillator is presented as follows:

$$\dot{\theta}_n(t) = \omega_n + \sum_{j=1}^{N}I_{nj}\omega_n\sum_{i=1}^{M}\frac{1}{2(2i-1)}\cdot(-1)^{(i+1)}\cdot\left[A_i\cos((2i-1)(\theta_n-\theta_j)) + B_i\sin((2i-1)(\theta_n-\theta_j))\right] \tag{9}$$

where $I_{nj} = s_{nj} \cdot I_0$ and $s_{nj}$ is the coupling strength.

Comparing the above with the equation in [1]:

$$\dot{\theta}_n(t) = \omega_n + \sum_{j=1}^{N}s_{nj}H(\theta_j - \theta_n) \tag{10}$$

We have the "averaged" connection function:

$$H(\theta_j - \theta_n) = \omega_n I_0 \sum_{i=1}^{M}\frac{1}{2(2i-1)}\cdot(-1)^{(i+1)}\cdot\left[A_i\cos((2i-1)(\theta_n-\theta_j)) + B_i\sin((2i-1)(\theta_n-\theta_j))\right] \tag{11}$$

where $I_0$ is the smallest unit of injection current in ONN and corresponds to the output current of one inverter in the coupler.

As can be seen above, if $\Gamma_n(t)$ is purely sinusoidal ($B_i=0$, M=1), then $H(\theta_j - \theta_n)$ will be sinusoidal as well and is easy to be analyzed. When it satisfies the condition: $s_{nj} = s_{jn}$ for all $n$ and $j$, the phase difference $\Delta\theta_n(t) = \theta_n(t) - \theta_1(t)$ is able to converge into a steady value, corresponding to a pattern [1].

## VI. PATTERN RECOGNITION SIMULATION OF MEMRISTOR-BASED ONN

Now we simulate the ONN for pattern recognition using the PPV of memristor-based oscillator obtained in section III.

The number of oscillators, N, is set to be 60 (corresponding to the pixels of 6*10) and that of patterns in storage, p, is set to be 3. The patterns stored in ONN are shown in Fig.8(upper part). Here black pixel corresponds to a phase difference of 0 rad while white pixel corresponds to π. We construct 2 groups of ONN. Group A: oscillators with sawtooth outputs (Fig.4 (a)). Group B: oscillators with sinusoidal outputs (in Fig.4(b)).

Case 1: The oscillators in ONN are chosen to be identical.Fig.8 (lower part) presents the results of recognition. Fig.9 shows the evolution of phase difference of group B in the process of initialization (0-0.7us) and recognition (0.7us-1.7us). In Fig.8, both groups are able to accomplish pattern recognition, but we find that (1) group B has more accuracy on the grayscale of picture, because the stabilized phase difference of the black pixel of group B (2.66 rad) is closer to its nominal value (π rad) than that of group A (2.356 rad); (2) group B needs less injecting current thus is more power-efficient than group A.

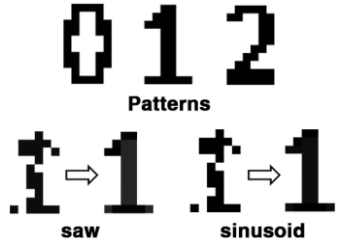

Fig.8 Patterns and Recognition in case 1

In fact, to ensure both groups to take the same time to initialize, the output current in group A is set to, $212uA$, which is 40 times larger than that in group B. The reason is that PPV/PRC value of group B (Fig. 4 b) is much larger than that of group A (Fig.4 a).

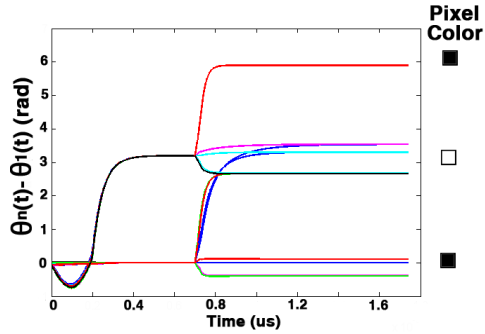

Fig.9 Phase evolution of each oscillator of group B in case 1

Case 2: To evaluate the robustness of ONN against frequency mismatch, we add ±10% frequency deviation for the oscillators in each group. As shown in Fig.10, after recognition, the pixels of group B (sinusoid) show more accurate recognition than group A(sawtooth).

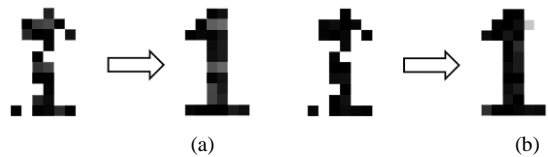

(a)          (b)

Fig.10 Recognition under frequency deviation(case 2) (a) Group A (sawtooth)   (b) Group B (sinusoid)

Finally, we compare the ONN simulation time of our method with that of other methods. After the process of initialization and recognition which lasts 30 us, all the simulations converge with correct pattern recognition. However, the simulation using the proposed method is at least 2000 times faster than that of the existing methods.

TABLE I SIMULATION TIME COMPARISON

| Method | Oscillator description | Coupler description | Simulation time |
|---|---|---|---|
| **Existing 1st** | Unfolding polynomial | Transistor level | 59 hours |
| **Existing 2nd** | Unfolding polynomial | Behavioral level | 6 hours |
| **Proposed** | **PPV** | Behavioral level | **9.967 sec** |

## VII. CONCLUSION

By abstracting PPV model of the memristor-based oscillator, we propose an efficient simulation methodology for the fast analysis of the large scale ONN synchronization and pattern recognition. Such method can apply to various memristor based oscillators, which allows for enormous speedup of simulation. It also reveals that both sinus and sawtooth output waveforms are capable of achieving pattern recognition, but appropriate adjustment of oscillator parameter with sinus output offers higher recognition accuracy and stronger mismatch immunity.